# Thermal Maps of Gases in Heterogeneous Reactions

Nanette N. Jarenwattananon[1], Stefan Glöggler[1], Trenton Otto[1], Arek Melkonian[1], William Morris[1], Scott R. Burt[2], Omar M. Yaghi[1,4,5†], Louis-S. Bouchard[1,3,4]

[1]Department of Chemistry and Biochemistry, [3]Biomedical Engineering Inter-Departmental Program, [4]California NanoSystems Institute, University of California Los Angeles, 607 Charles E. Young Dr. East, CA 90095-1569, USA.

[2]Department of Chemistry and Biochemistry, Brigham Young University, Provo, UT 84602

[†]Current address: Department of Chemistry, University of California, Berkeley, and The Molecular Foundry, Division of Materials Sciences, Lawrence Berkeley National Laboratory, Berkeley, California, 94720, United States.

[5]NanoCentury KAIST Institute and Graduate School of EEWS (WCU), Korea.

**Over 85% of all chemical industry products are made using catalysts (1), with the overwhelming majority of these employing heterogeneous catalysts (2) functioning at the gas-solid interface (3). Consequently, optimizing catalytic reactor design attracts much effort. Such optimization relies on heat transfer and fluid dynamics modeling coupled to surface reaction kinetics (4). The complexity of these systems demands many approximations, which can only be tested with experimental observations (5,6) of quantities such as temperature, pressure, concentrations, flow rates, etc. One essential measurement is a map of the spatial variation in temperature throughout the catalyst bed. We present here the first non-invasive maps of gas temperatures in catalyst-filled reactors, including high spatial resolution maps in microreactors enabled by parahydrogen. The thermal maps reveal energy flux patterns whose length scale correlates with the catalyst packing.**

**By exploiting the motional averaging under a weak applied magnetic-field gradient, the nuclear magnetic resonance (NMR) linewidths are inversely proportional to temperature. Measurements during the hydrogenation of propylene in reactors packed with metal nanoparticles and metal-organic framework catalysts yield temperature coefficients of ~0.1 Hz/K, and temperature error <4%. Temperature sensitivity increases with gradient strength, enabling tuning of precision.**

Catalysis is a fundamental component of many industrial processes, and considerable resources are spent optimizing these reactions. However, probing the behavior of gas inside a reactor is challenging because tools that do not disturb the flow are required. In addition to being non-invasive, such tools must be able to measure the physical parameters and gradients that dictate heat flow, mass flow, and catalytic efficiency (1-9). A critical parameter to understanding the thermodynamics of gas-solid reactions is gas temperature, not only because its accurate control is required for optimal operation, but also because temperature gradients in the reacting fluid reveal the energetics of a reaction in the presence of flow. Optical techniques have been applied to this problem (10-12) but their success is limited in opaque media. Sensor probe approaches based on thermocouples (13,14) or thin-film coated sensors (15) are utilized, but always lead to flow perturbations.

In this Letter, we show the successful use of NMR imaging and spectroscopy to generate thermal maps of gases inside gas-solid reactors. By applying a weak magnetic field gradient to broaden the spectral lines, we obtain a broadening that is temperature-dependent. Using this effect, we obtained, thermal maps with high spatial resolution (sub-millimeter) and accuracy (<4%). Thermal maps of hydrogenation reactions in catalyst beds containing platinum nanoparticles (PtNP) and metal-organic frameworks (MOFs) reveal that the flux of energy correlates with catalyst packing (section 3.5.3, *Supplementary Information*) as demonstrated on both small (10 mm) and micro (<1 mm) reactors. This technique provides a non-invasive tool to locate hot and cold spots in catalyst-packed reactors, offering unique capabilities for testing the approximations used in reactor modeling.

Traditional NMR thermometry is based on the temperature dependence of the chemical shift (16,17), spin-lattice and spin-spin relaxation times ($T_1$ and $T_2$, respectively) (18), diffusion in a pulsed-field gradient (PFG) (18), intermolecular multiple-quantum coherences (iMQC) (19), or contrast agents (20). However, these established methods are not practical for gas phase reactions: proton chemical shift and relaxation times ($T_1$, $T_2$) exhibit weak dependence on temperature (*vide infra*); signals from iMQC and PFG experiments are rapidly attenuated by diffusion; and contrast agents constitute an additive that may interfere with some reactions. In this Letter, we avoid these problems by using weak magnetic field gradients (<< 1 G/cm).

Our NMR technique is illustrated in Figure 1. Nuclear spins acquire a random phase shift due to the stochastic trajectories of molecules in the presence of a magnetic-field gradient. This dephasing leads to signal decay faster than $T_2$ and, consequently, spectral peaks are broadened beyond the natural linewidth (see also *Supplementary Information*). Higher temperatures accelerate molecular motion (21), leading to more efficient temporal averaging of the applied gradient, as illustrated in Fig. 1A. This method is a powerful complement to existing techniques for visualizing catalytic reactions under a range of experimental conditions (7-8, 22-29).

The pulse sequence (Figure S1, *Supplementary Information*) separates the phase-encoding scheme from the detection period to yield spectroscopic images. We apply a constant magnetic-field gradient during the detection period to introduce temperature-dependent line broadening. This gradient differs from a conventional frequency-encode gradient or PFG-NMR

(18) because it is much too weak to provide spatial encoding of the image or signal attenuation, respectively.

We tested this method over a temperature range of 293-443 K using a catalyst-packed reactor filled with catalyst supported on glass wool through which propylene and hydrogen reacted to form propane (see *Methods* section). This reactor is a laboratory-scale model reactor used to demonstrate the technique. The dependence of linewidth on temperature was calibrated over the entire reactor by omitting phase-encoding gradients and comparing the observed linewidth to temperature measured by fiber-optic temperature sensors. In the absence of an applied gradient, the chemical reactor system (Fig. 2) displays negligible temperature dependence (Fig. 1B). Thus, for this reactor, internal magnetic field inhomogeneities are too weak to produce significant temperature dependence of the linewidth. The situation changes, however, if we apply a gradient to deliberately broaden the lines. A modest gradient (0.10 G/cm) yields a temperature dependence of $(-0.130 \pm 0.006)$ Hz/K (Fig. 1B). Stronger gradients lead to stronger temperature dependence (Fig. 1B). The temperature dependence over this temperature range is well modeled by a linear regression. Sections 2 and 3 of the *Supplementary Information* discuss the robustness of this method relative to factors such as pressure, gas mixture composition, reactor type and reaction rate.

When the phase-encode gradients are used, Fourier transformation of the data set with respect to reciprocal space yields an image containing a 1D NMR spectrum in each voxel. The linewidth variations in each voxel are converted into temperatures using the above calibration. To get *absolute* temperatures, the temperature scale must be fixed by placing a sensor at a convenient location where perturbation of the flow is minimal (e.g., near an inlet or outlet).

We demonstrate thermal maps of two reactors, each packed with a different catalyst: Pt-nanoparticle (PtNP) or MTV-MOF-Pd (Pd-MOF) supported on glass wool. To bring the reaction into a regime of efficient conversion, the outer temperature of the reactor was maintained at 387 K using regulated external air flow. For the PtNP reactor system, a magnetic field gradient of 0.05 G/cm resulted in the following temperature dependence of the propylene linewidth:

$$\Delta f [Hz] = (-0.16 \pm 0.01) \left[\frac{Hz}{K}\right] \cdot T + (107 \pm 1)[Hz]. \quad \text{(eq. 1)}$$

For the Pd-MOF system and identical gradient, we observed the following temperature dependence:

$$\Delta f [Hz] = (-0.099 \pm 0.007) \left[\frac{Hz}{K}\right] \cdot T + (92 \pm 1)[Hz]. \quad \text{(eq. 2)}$$

The slopes of equations (1) and (2), together with the point sensor readout reference, yielded absolute temperature maps for the PtNP and Pd-MOF reactor beds, respectively (see also Section 3.5, *Supplementary Information*). Figure 3 shows the thermal images which have been thresholded at a signal-to-noise ratio (SNR) of 4 for axial, sagittal and coronal views. These thermal maps depict significant temperature variations throughout the catalyst bed, as would be expected for a heterogeneous reactor. These inhomogeneities correlate, in part, with the packing of the catalyst-impregnated glass wool (Fig. 3) and, in part, with the placement of the inlet and the outlet of the reactor. Due to non-uniformity in catalyst packing, some regions are more active than others, leading to differences in catalytic conversion ("hot spots" and "cold spots"). Another application of such thermal maps, if the heat capacity of the medium is known, would be to map spatial gradients in internal energy ($U$) of the reaction through the thermodynamic relationship $\nabla U = C_V \nabla T$. From these, we can derive a fundamental thermodynamic quantity, $\Delta U$, of the reaction, which reveals the flow of energy. The effective heat capacity of the medium ($C_V$) could potentially be inferred from micro-computed tomography (µ-CT) images such as those of Fig. 2. In section 3.5.3 (*Supplementary Information*) the characteristic length scales of $\|\nabla T\|$ are compared with those of the µ-CT map in similar reactors.

To validate the NMR-derived temperature measurements *in situ*, three fiber optic temperature readouts were made at the center of the field of view (FOV), one 4 mm below, and one 4 mm above (Fig. 3). The sensors in the PtNP reactor measured temperatures of 398 K, 412 K, and 427 K, which correspond well to the NMR-derived temperatures in the XZ map of 400 K, 417 K, and 424 K. Similarly, the temperatures in the Pd-MOF reactor were measured as 382 K, 411 K, and 425 K versus the NMR-derived measurements in the YZ map of 389 K, 416 K, and 441 K. The random error in the NMR-derived temperature measurement,

$$\frac{\delta T}{T} = \sqrt{\left(\frac{\delta(\Delta f)}{(\Delta f)}\right)^2 + \left(\frac{\delta(\Delta f/\Delta T)}{(\Delta f/\Delta T)}\right)^2} \qquad \text{(eq. 3)}$$

was calculated to be 4% on the Kelvin scale using the relative errors in the linear regression, $\frac{\delta(\Delta f)}{(\Delta f)}$ and $\frac{\delta(\Delta f/\Delta T)}{(\Delta f/\Delta T)}$, where *Δf* is the linewidth from the fit of the spectrum and *Δf/ΔT* is the slope of the temperature calibration curve. The disagreement between NMR-derived temperatures and fiber optic sensor measurements was at most 4%; thus, systematic errors for the systems studied were not significant. The correspondence of these results demonstrates the ability to map temperatures of reacting gases inside an operating catalytic reactor with millimeter resolution using NMR signal from thermally polarized protons.

The 10-mm reactor in the above experiments was chosen for validation purposes: 1-mm diameter point-sensors cause minimal perturbation of the flow because they are much smaller than the reactor itself. With the method validated, smaller reactors can be investigated with a single reference measurement located at a convenient location *outside* the reactor bed to avoid perturbing the flow. Thermal maps of a microreactor are shown in Figure 4, where *para*-state enriched hydrogen (24-30) was used as reactant to overcome the loss of signal associated with the smaller image voxels. This reactor is packed without glass wool support. Results on additional catalyst-packed reactors are presented in Section 3.4 (*Supplementary Information*).

We argued earlier that NMR thermometry based on $T_1$ or $T_2$ is either impractical or insensitive when applied to gas-phase reactions. To verify this, $T_2$ was measured over the temperature range 303-413 K (see *Methods*); it did not show statistically significant temperature dependence. Inversion recovery measurements showed that $T_1$ is temperature dependent (-2.3±0.2 ms/K). However, these experiments required a minimum of 51 inversion times to yield temperature measurements with comparable precision; using this approach to create thermal maps would require a scan time of 1,000 hours. If the number of inversion-recovery steps is reduced to four, the error in temperature a factor of forty greater than our motional-averaging technique and the scan time is still five times longer (~2.5 hrs) than our technique (~30 min).

The motional-averaging NMR method reported here opens the door to studies of *in situ* thermodynamics and optimization of gas-phase reactors. Our technique outperforms other existing thermometry techniques, does not require added contrast agents, is non-invasive and can provide detailed views of spatial variations in temperature. Another advantage is that the temperature sensitivity is tunable: the stronger the applied gradient, the stronger the dependence on temperature and the higher the sensitivity of the method. In practice, the gradient should be significantly stronger than the internal magnetic field inhomogeneities to ensure that the temperature calibration coefficient is essentially sample-independent, but weak enough to avoid overlap of spectral lines. Good results are obtained by broadening the resonances by factors of 5 to 10 (see Section 3.5.2, *Supplementary Information* for broadening across voxel vs. reactor). This method can be extended to mixed (gas-liquid-solid) reactions provided at least one gas-phase species contains hydrogen nuclei. The method is largely independent of gas mixture composition, as discussed in Sections 2.2, 3.2 and 3.3 (*Supplementary Information*), and can operate across a wide range of temperatures, pressures (see Sections 2 and 3.2, *Supplementary Information*) and reactor types (Section 3.4, *Supplementary Information*). The temperature reported is that of the gaseous component contributing the NMR signal, which in many cases, is in local thermal equilibrium with the solid phase (Section 2.3, *Supplementary Information*).

**Methods Summary**

NMR spectroscopy and imaging were performed using a Varian 9.4 T VNMRS 400 MHz magnet with air-regulated VT control and micro-imaging gradients. Samples were allowed to thermally equilibrate for 10 minutes at each temperature followed by probe retuning. The temperature calibration was performed using FISO FOT-L fiber optic temperature sensors (FISO Technologies, Inc., Québec, Canada). Propylene and hydrogen gases (ultra high purity) were obtained from Airgas. Data was processed using VnmrJ 2.3 A and MATLAB (The MathWorks, Natick, MA).

The catalysts were prepared according to reported procedures (29-32). Details about the motional-averaging technique calibration, temperature mapping, $T_1$ and $T_2$ relaxometry, and microreactor imaging (30), can be found in Methods.

**Supplementary Information** is linked to the online version of the paper at www.nature.com/nature.

**Acknowledgements** We thank Prof. J. Reimer for useful discussions; Profs. N. K. Garg, W. Gelbart, and C. Knobler for reading the manuscript; M.T. Yeung for technical help with MATLAB; J. Brown and R. Sharma for assistance with chemical synthesis. This work was partially funded by Dreyfus and NSF CHE-1153159 grants (L.S.B.), and BASF (Ludwigshafen, Germany) for synthesis and DOE for porosity measurements (O.M.Y.).

**Author Information** Reprints and permissions information is available at www.nature.com/reprints. The authors declare no competing financial interests. Correspondence should be addressed to L.S.B. (bouchard@chem.ucla.edu).

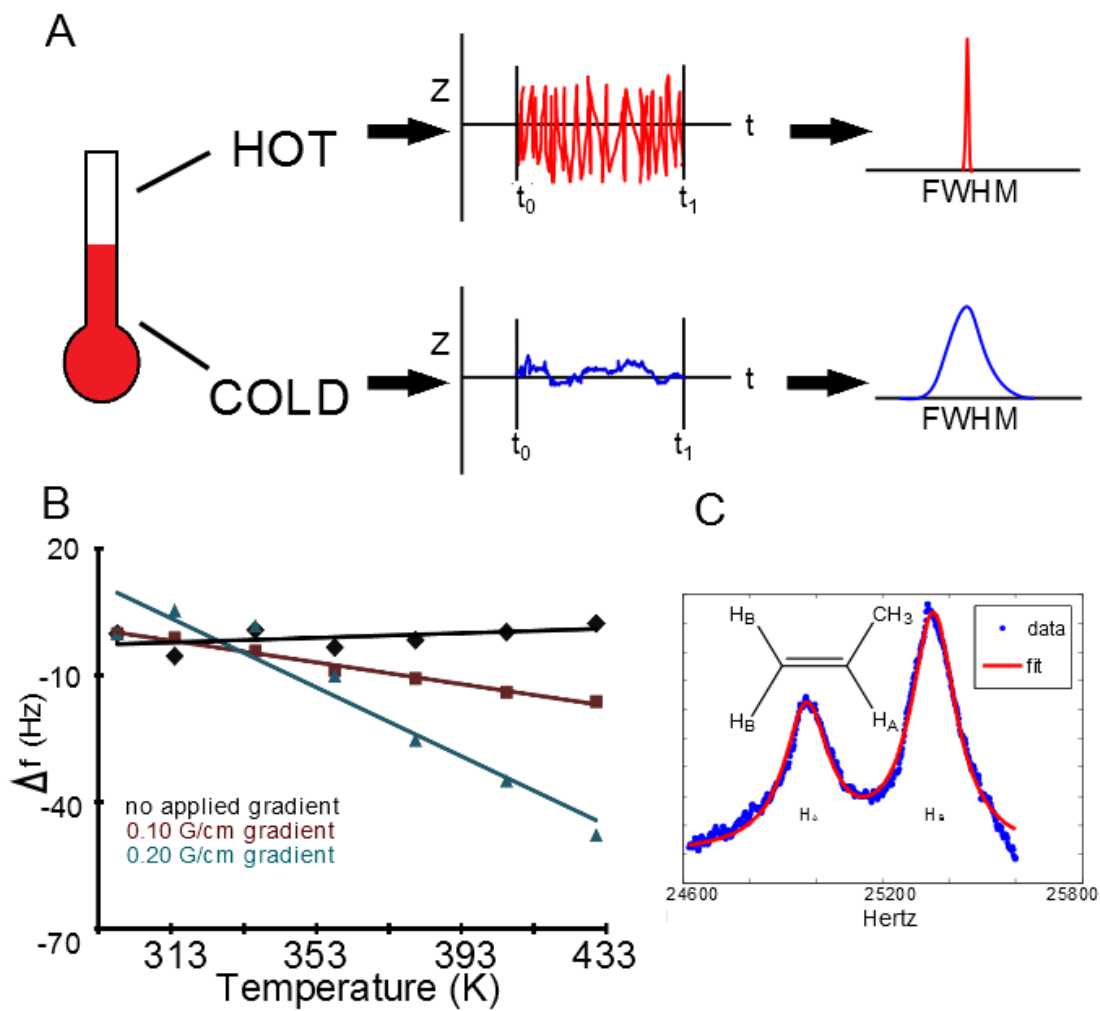

**Fig. 1.** Illustration of the motional-averaging method. **A** At higher temperatures, the gas molecules move faster, with motion averaging the effects of the dephasing gradient. This motional averaging leads to narrow linewidths at higher temperatures. **B** Without a gradient, the linewidth is determined by homogeneous and inhomogeneous broadening mechanisms, which generally show a weak dependence on temperature. If we apply an external gradient, the dependence of linewidth on temperature is much larger and independent of the nature of the porous medium. **C** For the calibration experiment, we fit the propylene spectrum at each temperature increment to a sum of Lorentzian curves using nonlinear least squares.

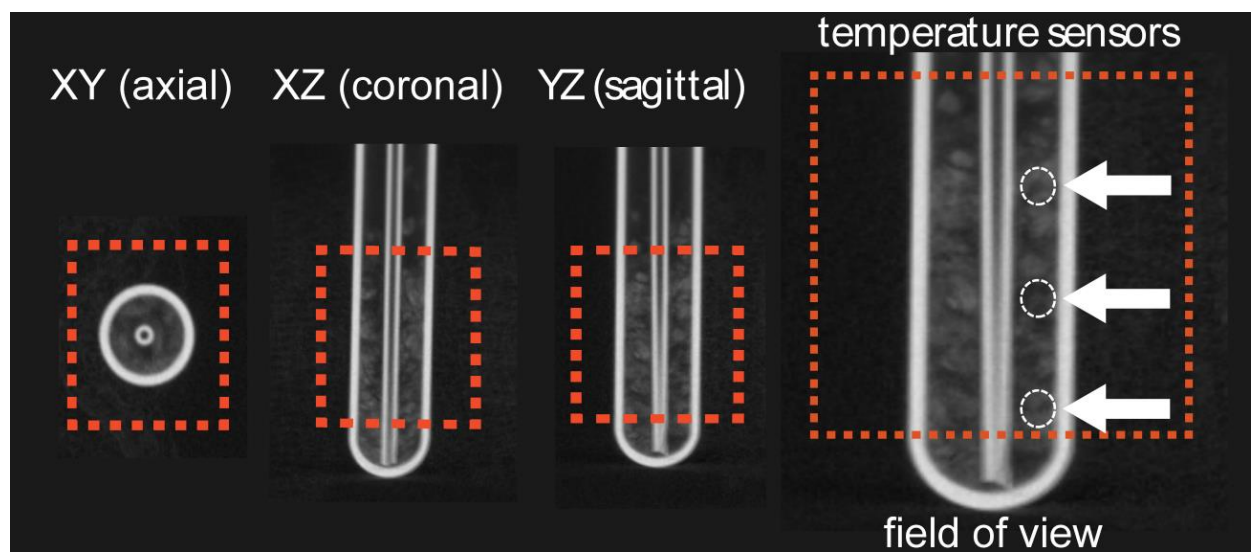

**Fig. 2.** Orthogonal slices of the chemical reactor system by micro-computed tomography (μ-CT). The reactor consists of a 10-mm pyrex NMR tube, a gas inlet tube, and a heterogeneous catalyst loaded on glass wool. The glass wool is depicted by greyscale patterns. Temperature measurements were taken with fiber-optic probes at three different locations (indicated by arrows) within the NMR field of view (FOV) to verify NMR-derived temperature measurements. The approximate location of the FOV is indicated by the dashed box. The axial, coronal and sagittal planes of the μ-CT images correspond to those of Figure 3.

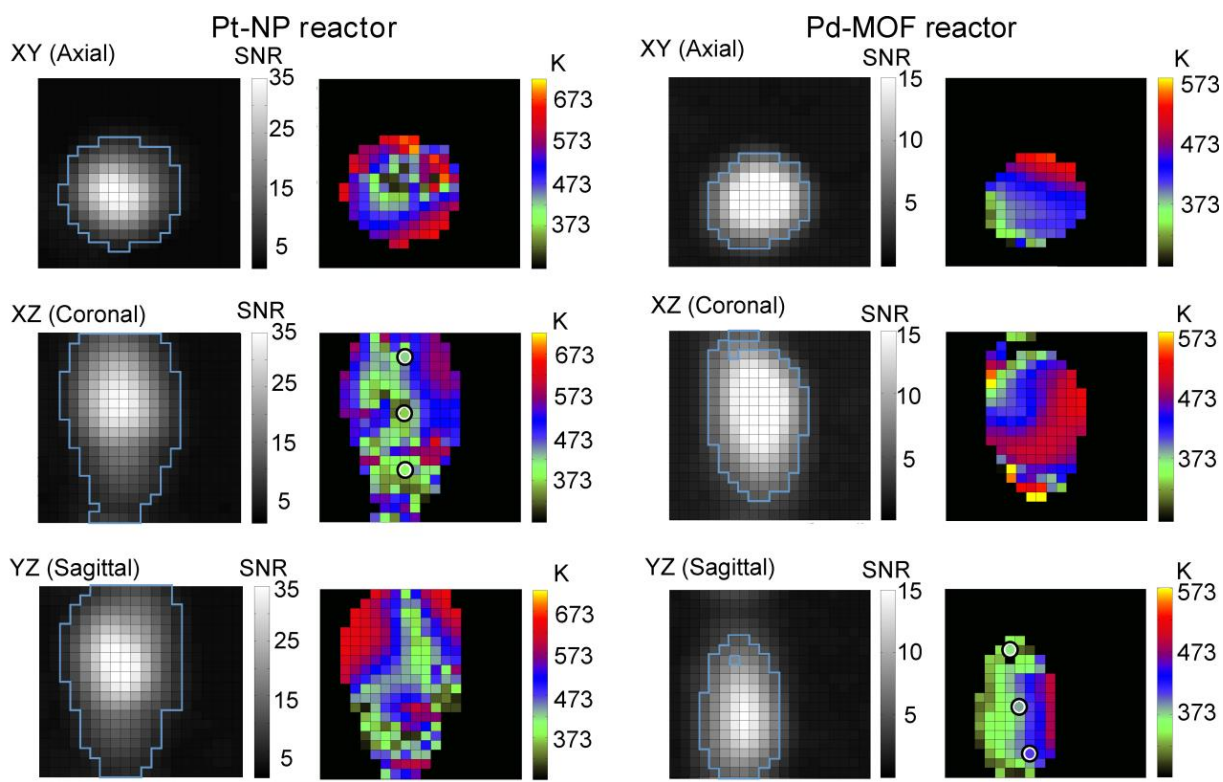

**Fig. 3.** 2D NMR signal-to-noise ratio (SNR) maps (greyscale) and the corresponding temperature maps (color) of the 10 mm Pt-nanoparticles (PtNP) and MTV-MOF-Pd (Pd-MOF) reactor systems. The SNR images detail the signal intensity of proton resonances used to derive the temperature; pixels with an SNR greater than four are enclosed by the blue outline. Each pixel corresponds to a region 0.73×0.73 mm$^2$ in-plane and is assigned a temperature based on the linewidth of the spectrum acquired. Temperature was only calculated for pixels that demonstrated SNR greater than four. The temperature measurement was confirmed by independent fiber-optic thermometry (FISO FOT-L sensor) with temperature sensors placed within the FOV, as indicated by circles in the thermal maps. The alignment of the axial, sagittal and coronal planes in relation to the reactor is shown in Figure 2.

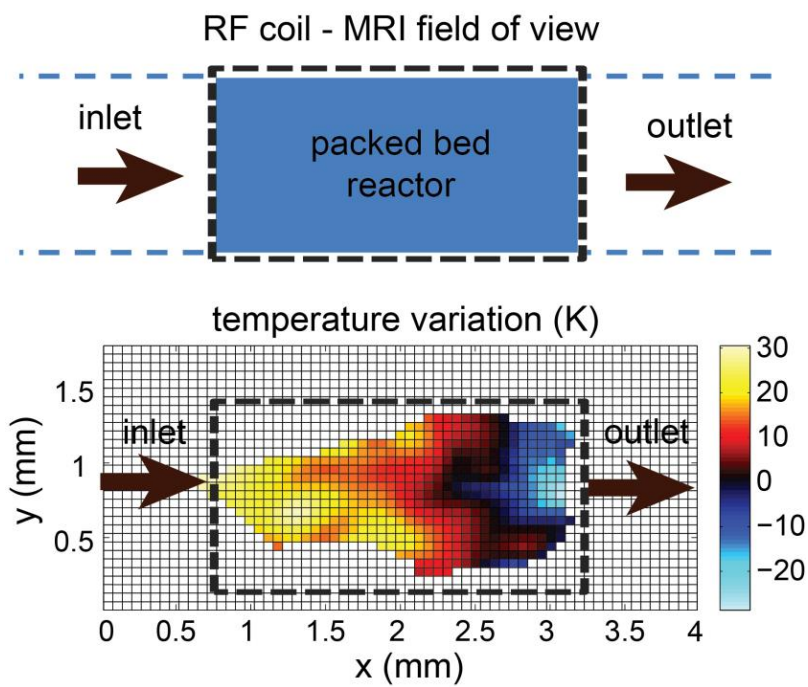

**Fig. 4.** Temperature map in a catalytic microreactor. Signal enhancement from parahydrogen was used to provide detailed views inside the microreactor. In this experiment, the reactor was held at approximately 418 K by using a variable temperature controller to preheat the gas flowing through a heat exchange coil. The temperature calibration coefficient was -0.200±0.006 Hz/K. The thermal map depicts temperature changes relative to the incoming gas.

**Methods**

**Pt-Nanoparticles Catalyst Preparation**: the synthesis of the Pt-nanoparticles follows the protocol published by Sharma and Bouchard (29). The supported heterogeneous catalyst used in reaction imaging was made by soaking a piece of glass wool ($SiO_2$) in a solution of *p*-mercaptobenzoic-capped (2.5 ± 0.4) nm Pt-nanoparticles to yield 1-wt% $Pt/SiO_2$.

**MTV-MOF-Pd Catalyst Preparation**: MTV-MOF-AB was prepared in accordance with reported procedures (31). The free amine ($-NH_2$) was metallated with palladium, by postmodification to give MTV-MOF-Pd (32). The MTV-MOF-Pd catalyst was then mixed with $TiO_2$ nanopowder and deposited on glass wool ($SiO_2$) to obtain the supported heterogeneous catalyst used in reaction imaging.

**Calibration of Motional-Averaging Technique**: A 10 mm NMR tube was packed with catalyst-loaded glass wool. For flowing propylene gas (40 PSI, 15 cc/min), a FID was acquired between 303 K and 413 K in steps of 10 K. A spin-echo pulse sequence was used, and a magnetic field gradient was applied during acquisition. Spectral windows were fit to a sum of Lorentzian lines with MATLAB using nonlinear least-squares.

**Temperature Mapping**: Propylene gas (40 PSI, 15 cc/min) and parahydrogen gas (40 PSI, 15 cc/min) were flowed through the system. Each image was acquired using a spin-echo imaging pulse sequence with 441 transients during a continuous flow reaction for a total acquisition time of 33 minutes. The scan time, which enables images from a steady-state flow, could be further reduced by compressed sensing. The spectrum of each pixel containing a signal-to-noise ratio (SNR) greater than four was analyzed in MATLAB, using nonlinear least-squares to fit the propane and propylene resonances to a sum of Lorentzian curves. The temperature measurement was confirmed by a FISO FOT-L sensor. The spatial resolution of the experiment is limited by the SNR of the image, as in conventional MRI experiments. For higher spatial resolutions (smaller voxels), the applied gradient can be even weaker. This is because the gradient only needs to be about 5-10 times larger than the linewidth over the voxel in the absence of a gradient.

*$T_1$ Relaxometry*: A 10 mm NMR tube with a J. Young Valve adaptor was pressurized to 15 PSI with propylene gas. An inversion recovery pulse sequence with 51 inversion times was applied, and relaxation time was determined by fitting the data to a non-linear least squares exponential recovery curve in MATLAB. $T_1$ relaxation time was determined between 303 K and 413 K in steps of 10 K.

*$T_2$ Relaxometry*: A 10 mm NMR tube with a J. Young Valve adaptor was pressurized to 15 PSI with propylene gas. A single shot Carr-Purcell-Meiboom-Gill (CPMG) echo train of 100 echoes was applied for interpulse spacing between 1-20 ms. Relaxation times were determined by fitting the amplitude of alternating echoes to a non-linear least squares exponential decay curve in MATLAB. $T_2$ relaxation time was determined between 303 K and 413 K in steps of 10 K.

**Microreactor Imaging**: The catalyst, Wilkinson's catalyst supported on silica gel, and microreactor were prepared as described in (30). Parahydrogen induced polarization, PHIP (with hydrogen 50% *para*-enriched), was used to provide signal enhancement from the microreaction. The calibration and data analysis were done similarly as described above.